\documentclass[amsmath,amssym,aps,prl,reprint]{revtex4-1}
\usepackage{lmodern}
\usepackage{amssymb,amsmath}
\usepackage{ifxetex,ifluatex}
\usepackage{fixltx2e} 
\ifnum 0\ifxetex 1\fi\ifluatex 1\fi=0 
  \usepackage[T1]{fontenc}
  \usepackage[utf8]{inputenc}
\else 
  \ifxetex
    \usepackage{mathspec}
  \else
    \usepackage{fontspec}
  \fi
  \defaultfontfeatures{Ligatures=TeX,Scale=MatchLowercase}
\fi
\IfFileExists{upquote.sty}{\usepackage{upquote}}{}
\IfFileExists{microtype.sty}{%
\usepackage{microtype}
\UseMicrotypeSet[protrusion]{basicmath} 
}{}

\usepackage[unicode=true]{hyperref}
\hypersetup{
            pdftitle={Noise and diffusion of a vibrated self-propelled granular particle},
            pdfborder={0 0 0},
            breaklinks=true}
\usepackage{graphicx,grffile}
\makeatletter
\def\maxwidth{\ifdim\Gin@nat@width>\linewidth\linewidth\else\Gin@nat@width\fi}
\def\maxheight{\ifdim\Gin@nat@height>\textheight\textheight\else\Gin@nat@height\fi}
\makeatother
\setkeys{Gin}{width=\maxwidth,height=\maxheight,keepaspectratio}
\IfFileExists{parskip.sty}{%
\usepackage{parskip}
}{
\setlength{\parindent}{0pt}
\setlength{\parskip}{6pt plus 2pt minus 1pt}
}
\renewcommand{\section}[1]{}
\renewcommand{\subsection}[1]{}
\renewcommand{\subsubsection}[1]{}
\ifx\paragraph\undefined\else
\let\oldparagraph\paragraph
\renewcommand{\paragraph}[1]{\oldparagraph{#1}\mbox{}}
\fi
\ifx\subparagraph\undefined\else
\let\oldsubparagraph\subparagraph
\renewcommand{\subparagraph}[1]{\oldsubparagraph{#1}\mbox{}}
\fi

\makeatletter
\def\fps@figure{htbp}
\makeatother

\newcommand{\angles}[1]{\left\langle #1 \right\rangle}

\begin{document}
\title{Noise and diffusion of a vibrated self-propelled granular particle}

\author{Lee Walsh}
\email{lawalsh@physics.umass.edu}
\affiliation{Department of Physics, University of Massachusetts, Amherst, USA}

\author{Caleb G. Wagner}
\affiliation{Martin Fisher School of Physics, Brandeis University, Waltham, MA, USA}

\author{Sarah Schlossberg}
\affiliation{Department of Physics, University of Massachusetts, Amherst, USA}

\author{Christopher Olson}
\affiliation{Department of Physics, University of Massachusetts, Amherst, USA}

\author{Aparna Baskaran}
\affiliation{Martin Fisher School of Physics, Brandeis University, Waltham, MA, USA}

\author{Narayanan Menon}
\email{menon@physics.umass.edu}
\affiliation{Department of Physics, University of Massachusetts, Amherst, USA}

\date{\today}
\begin{abstract}
Granular materials are an important physical realization of active
matter. In vibration-fluidized granular matter, both diffusion and
self-propulsion derive from the same collisional forcing, unlike many
other active systems where there is a clean separation between the
origin of single-particle mobility and the coupling to noise. Here we
present experimental studies of single-particle motion in a vibrated
granular monolayer, along with theoretical analysis that compares grain
motion at short and long time scales to the assumptions and predictions,
respectively, of the active Brownian particle (ABP) model. The results
demonstrate that despite the unique relation between noise and
propulsion, granular media do show the generic features predicted by the
ABP model and indicate that this is a valid framework to predict
collective phenomena. Additionally, our scheme of analysis for
validating the inputs and outputs of the model can be applied to other
granular and non-granular systems.
\end{abstract}
\maketitle

\section{Introduction}\label{introduction}

\subsection{Background}\label{background}

Active
materials~\citep{ramaswamy_mechanics_2010, marchetti_hydrodynamics_2013}
manifest striking nonequilibrium collective dynamics because they are
made up of self-propelled entities. That is, in addition to diffusive
motion produced by a noisy environment, their constituent particles also
have propulsion along some internal, body-fixed axis. Examples include
flocks of creatures on land and in the sea and sky; cells and \emph{in
vitro} cell extracts; and phoretic colloids and emulsions. Some of the
most striking experimental demonstrations of nonequilibrium collective
effects were first shown in vibrated granular matter. These include
spontaneous vortex formation~\citep{blair_vortices_2003}, coupling of
orientability and activity to produce persistent flows and giant number
fluctuations~\citep{narayan_long-lived_2007}, boundary
migration~\citep{deseigne_vibrated_2012}, and modified crystallization
dynamics~\citep{briand_crystallization_2016}.

\subsection{Motivation}\label{motivation}

These fascinating collective effects can theoretically be shown to
emerge from microscopic models in which there is an abstracted
description of single-particle motion. A theoretical cornerstone is the
active Brownian particle (ABP) model~\citep{marchetti_minimal_2016}, in
which a Brownian particle propels itself with a constant speed along a
body-fixed axis that rotates diffusively~\footnote{A related but
  distinct model is the run-and-tumble
  model~\citep{tailleur_statistical_2008, cates_when_2013}.}. The two
key dynamical ingredients in this model---self-propulsion and
diffusion---are independent physical processes that together contribute
to the single-particle motion. The assumptions of this popular model are
well satisfied by some physical systems, and it has successfully
predicted, among other phenomena, phase separation and an active solid
phase in self-propelled
particles~\citep{fily_athermal_2012, redner_structure_2013, stenhammar_continuum_2013, wysocki_cooperative_2014, cates_motility-induced_2015, redner_classical_2016, krinninger_nonequilibrium_2016}.

Self-propulsion in granular particles arises from different physical
considerations than in other soft and living systems. Here, both the
propulsion and the noise share the same non-thermal origin. In a
two-dimensional granular fluid, dynamics are driven by a vibrating
boundary which energizes particles via collisions in the vertical
direction. These collisions act as a source of high-frequency noise that
leads to diffusion in the other two
dimensions~\citep{yadav_diffusion_2012}. Particles with anisotropy in
shape~\citep{tsai_chiral_2005, narayan_nonequilibrium_2006, daniels_dynamics_2009}
or other properties~\citep{deseigne_vibrated_2012} will also be
preferentially propelled along a body-fixed axis by the collisions.
Further, collisions are locked to the deterministic vibrations of the
plate. This could lead to temporal or spatial correlations of the noise,
as well as a non-thermal noise
spectrum~\citep{zik_mobility_1992, danna_observing_2003}.

\subsection{Goal}\label{goal}

In this article, we seek to quantify single-particle motion of a
vibrated granular system and to test the theoretical paradigm in this
setting. To do so, we measure the noise-driven short-time motion and
check the assumptions of the ABP model. We then compare the mean
long-time motion with the predictions of the model. We show that model
parameters derived from the short- and long-time measurements are
self-consistent. To test the robustness of our results, we have carried
out this procedure on various combinations of particle design,
containment, and vibration. More generally, we present a template for
systematically relating observed short- and long-time motion to each
other and to the predictions of the active Brownian particle model, not
just in granular matter but in other active particle systems as well.

\subsection{Method}\label{method}

In the simplest version of the ABP model, the center-of-mass velocity of
an object is the sum of two contributions, a constant velocity~\(v_0\)
directed along a body-fixed orientation
\(\hat n = {\cos \theta \choose \sin \theta}\), and a translational
noise~\(\vec \eta\). This is accompanied by rotation of \(\hat n\)
driven by a noise term~\(\xi\).
\begin{align}
\dot{\vec{r}}(t) &= v_0 \hat n(t) + \vec \eta(t) \label{eq:abp_T} \\
\dot{\theta}(t) &= \xi(t). \label{eq:abp_R}
\end{align}
The translational and rotational noise terms are Gaussian with zero
mean, no spatial or temporal correlations, and variances \(2D_T\) and
\(2D_R\), respectively. Thus the short-time dynamics are fully
characterized by three parameters, \(D_R\), \(D_T\), and \(v_0\), which
determine the statistical behaviour of the particle over longer times.
An equivalent parameterization can be given in terms of the two
diffusion coefficients and a persistence length~\(\ell_p = v_0/D_R\).

\begin{figure}
\includegraphics[width=\columnwidth]{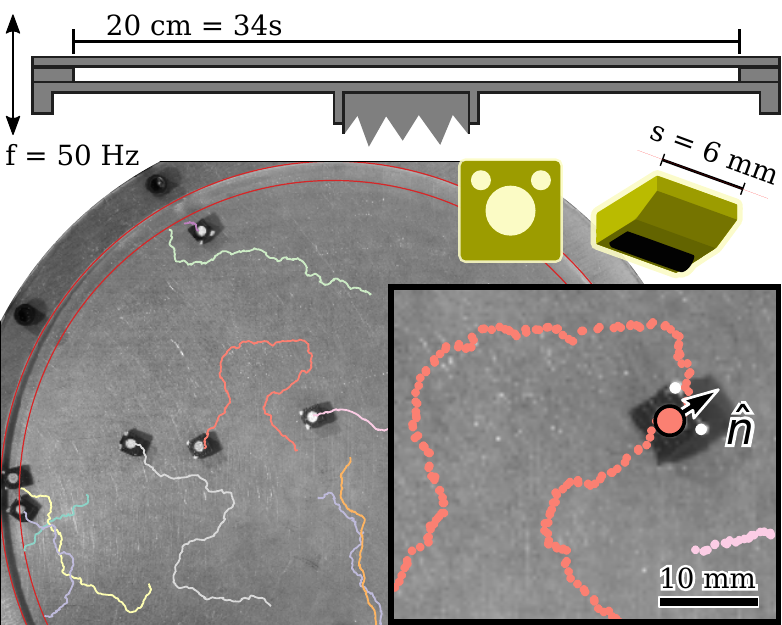}
\caption{
    A diagram of the experimental apparatus, showing a side view of the
    vibrating chamber and top and oblique views of a particle. The lower
    panel shows a video frame overlaid with the tracked positions of
    each particle for a segment of time of about 80 vibration periods.
    The lower right panel zooms in on a track to show individually
    detected positions $\vec r(t)$ and the propulsion axis $\hat n(t)$.
}
\label{fig:diagram}
\end{figure}

\section{Experiment}\label{experiment}

\subsection{System}\label{system}

\subsubsection{Introduce experiment.}\label{introduce-experiment.}

\subsubsection{Shaking apparatus to limit any non-vertical
motion.}\label{shaking-apparatus-to-limit-any-non-vertical-motion.}

\emph{Experiment.---} The self-propelled grains in our experiment are
confined in a cell with a circular aluminium base (of diameter 203~mm)
and an acrylic lid, as shown in
figure~\ref{fig:diagram}~\citep{walsh_ordering_2016}. The cell is
vibrated vertically by an electromagnetic shaker (LDS456) through a
flexible coupling to a square air bearing that constrains horizontal
motion (see for example \citep{harris_generating_2015}). The vibration
frequency in these experiments is held fixed at \(f = 50\)~Hz, and the
vibration amplitude is varied from 10 to \(20g\).

\subsubsection{Tracking procedure.}\label{tracking-procedure.}

Particle motions are captured by video imaging at a frame rate of
120~fps \(=2.4f\). Every particle is marked with a central white dot to
detect position \(\vec{r}\) with a precision of \(0.02s\); two smaller
dots at the corner define the orientation \(\theta\) of the propulsion
axis with a precision of 0.03 rad. The positions and orientations are
tracked over time to generate individual particle trajectories.

\subsubsection{Describe experiment -- system size, particle
size.}\label{describe-experiment-system-size-particle-size.}

All particles we use have a square footprint of width
\(s = 6.28 \pm 0.04\)~mm, a maximum height of \(3.9 \pm 0.1\) mm, and
are made of ABS thermosetting resin. The cell height of 4.76~mm
constrains the particles to quasi-two-dimensional motion. Since our goal
in this article is to study single-particle motion, we maintain a low
area fraction of particles, (\(\phi \sim 2\%\)) so that collisions
between particles are rare. Particles within \(1.5s\) of the cell
boundary are excluded from the analysis to eliminate any edge effects.

\subsection{Data}\label{data}

\subsubsection{Energy and momentum injection are determined by
particle-wall
interaction.}\label{energy-and-momentum-injection-are-determined-by-particle-wall-interaction.}

Asymmetry in the particle design causes them to behave as self-propelled
polar `walkers' by rectifying the noise provided by the vertical
vibration (see figure~\ref{fig:diagram}). Dynamical asymmetry is
introduced both through the particle geometry as well as by frictional
properties~\citep{narayan_long-lived_2007, kudrolli_concentration_2010, deseigne_vibrated_2012}:
a short length of nitrile rubber inset along one edge provides enhanced
friction, while a bevel on the opposite edge produces asymmetric rocking
of the particle following a collision with the
floor~\citep{supplementary}. Tuning the design parameters, along with
the confining gap and the vibration amplitude, provides some control
over the parameters of motion. In this article, we exploit this
flexibility to test the quantitative predictions of the active Brownian
particle model for six distinct configurations~\citep{supplementary} of
walkers. We first discuss the short-time dynamics, followed by the
long-time mean behavior, showing data from one of these configurations.

\begin{figure}
\includegraphics[width=\columnwidth]{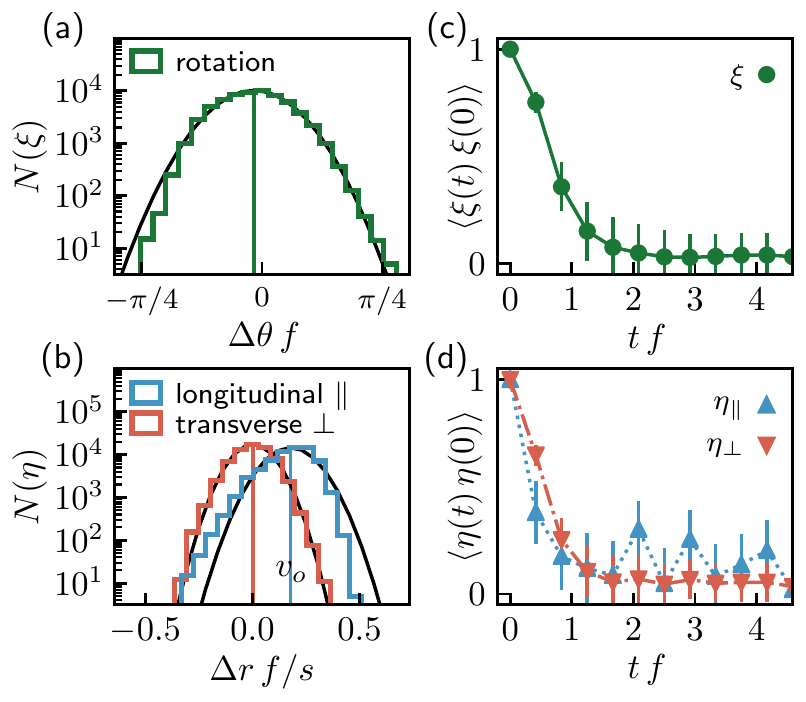}
\caption{
    The noise is characterized by the distribution of short-time
    (a)~rotational $\xi$ and (b)~translational displacements, the latter
    resolved parallel $\eta_\parallel$ and perpendicular $\eta_\perp$ to
    the mobility axis $\hat n$. When compared with the Gaussian
    distribution (solid line) assumed in the minimal ABP model, $N(\xi)$
    and $N(\eta_\parallel)$ show some skewness. The mean of
    $\eta_\parallel$ determines the model parameter $v_0$, and the
    variances of $\xi$ and $\eta$ are used to estimate $D_R$ and $D_T$.
    The rapid decay of autocorrelations (c)~and (d)~of the noise
    components supports the model assumption that the noise is
    $\delta$-correlated.
}
\label{fig:velocity}
\end{figure}

\begin{figure*}
\includegraphics[width=\textwidth]{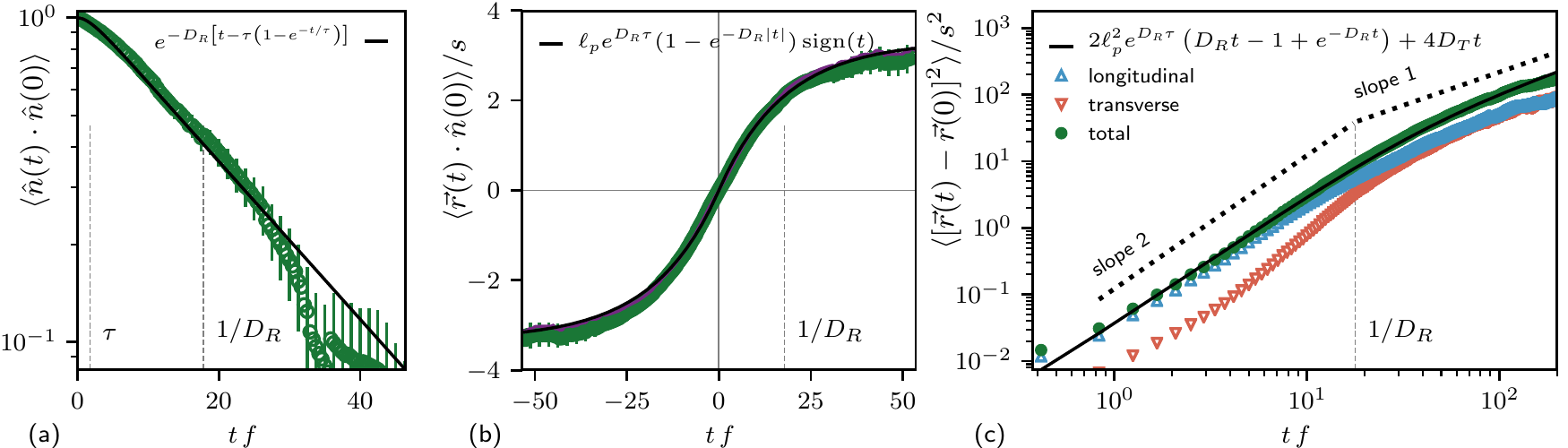}
\caption{
    Orientation--displacement correlation functions fitted by the ABP
    predictions. (a) Orientation autocorrelation $\angles{\hat n(t)
    \cdot \hat n(0)}$ shows the persistence of the particle orientation
    over time; the fit yields $D_R$. (b) Longitudinal displacement
    $\angles{\vec r(t) \cdot \hat n(0)}$ shows the mean displacement
    along the orientation at $t = 0$; the fit yields $\ell_p$. (c) Mean
    squared displacement $\angles{\left[\vec r(t) - \vec r(0)
    \right]^2}$; the fit yields $D_T$. We also resolve the m.s.d. in the
    body-frame to show the longitudinal and transverse components of the
    displacement.
}
\label{fig:correlations}
\end{figure*}

\section{Noise in vibrated particles}\label{noise-in-vibrated-particles}

\subsection{Direct noise characterization (short-time
dynamics)}\label{direct-noise-characterization-short-time-dynamics}

\emph{Characterization of noise.---} To directly characterize the noise
generated by particle-wall collisions, we measure the translational and
rotational velocity at a short timescale. Recall that the ABP model
assumes that the noise terms are Gaussian-distributed and uncorrelated
over space and time. These velocity components are calculated from the
numerical derivative of the orientation and position of the
particle~\citep{supplementary}. The translational velocity is resolved
into a longitudinal~\(\dot r_\parallel\) and a
transverse~\(\dot r_\perp\) component, parallel and perpendicular
to~\(\hat n\), respectively.

\subsubsection{Velocity distribution}\label{velocity-distribution}

Histograms for the velocities are shown in figure~\ref{fig:velocity}.
The distributions all show a single peak, and no long tails. The moments
from these distributions characterize the noise parameters: the
variances provide values for the diffusion constants, and the mean of
the longitudinal velocity is the short-time measurement for~\(v_0\).
Higher moments show small deviations from a pure Gaussian distribution,
with perhaps the largest quantitative departure being the skewness of
the longitudinal velocity component~\citep{supplementary}.

\subsubsection{Time autocorrelation}\label{time-autocorrelation}

The autocorrelations of the rotational and translational velocities show
only very short time correlations~(see figure~\ref{fig:velocity}). The
longest is that of the rotational autocorrelation, whose decay time
\(\tau\) is on the order of the vibration period~\(1/f\). We incorporate
this small but finite correlation time into the ABP
model~\citep{daniels_dynamics_2009} as an exponential correlation in the
rotational noise
\(\angles{\xi(t) \, \xi(t')} = (D_R/\tau)e^{-|t-t'|/\tau}\)~\citep{supplementary}.

\subsubsection{Spatial correlations}\label{spatial-correlations}

Spatial noise correlations between particles could lead to more complex
collective behaviors not accounted for by minimal models such as the
ABP. Despite the highly correlated nature of the noise source over space
as well as time, we find no significant spatial correlations between
particles. We compute interparticle spatial correlations as a function
of particle separation distance, which do not persist beyond steric
interactions~\citep{supplementary}. Thus the nonrandom and highly
correlated motion of the plate does not induce correlations between
particles.

\subsection{Mean particle behavior (long-time
correlations)}\label{mean-particle-behavior-long-time-correlations}

\begin{figure*}
\includegraphics[width=\textwidth]{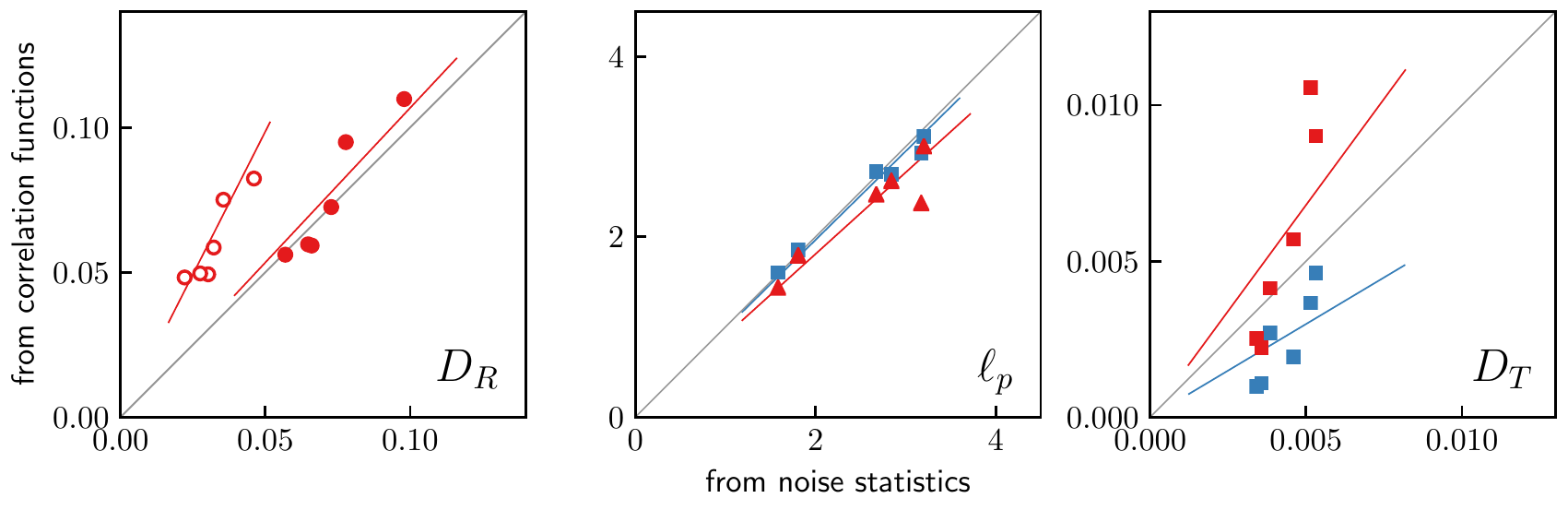}
\caption{
    A comparison of dynamical parameters obtained from short-time noise
    measurement (horizontal axis), against the long-time values obtained
    from fitting to the correlation functions (vertical axis). There is
    good agreement for the rotational diffusion constant~$D_R$ (left
    pane) and persistence length $\ell_p = v_0/D_R$ (center), while the
    translational diffusion constant~$D_T$ (right) is poorly determined.
    Each of the six data points corresponds to a different experimental
    configuration~\citep{supplementary}, while different markers
    distinguish among fits. Color indicates number of free parameters:
    red markers are single-parameter fits, blue are two-parameter fits.
    Shape indicates the correlation function being fit: circles
    ($\bullet$) are fits to the orientation autocorrelation in
    figure~\ref{fig:correlations}(a); triangles ($\blacktriangle$) are
    fits to the displacement--orientation correlation in
    figure~\ref{fig:correlations}(b); squares ({\tiny $\blacksquare$})
    are fits to the mean squared displacement in
    figure~\ref{fig:correlations}(c). Open symbols ($\circ$) indicate
    we have neglected correlated rotational noise (with
    timescale~$\tau$).
}
\label{fig:parametric}
\end{figure*}

\emph{Long-time diffusion and mobility.---} We turn to characterizing
the longer-term motions of the particles, via three correlation
functions built from the observed~\(\vec r(t)\) and~\(\hat n(t)\). These
data are compared to functions obtained from integrating the active
Brownian particle model over time and averaging over the noise
distribution. Some of these or similar functions have been measured
before~\citep{yadav_diffusion_2012, ten_hagen_brownian_2011}, but here
we perform a sequence of fits to each one of the three pertinent ABP
model parameters. We show correlation functions and fits in
figure~\ref{fig:correlations} for one of our experimental
configurations.

We first examine the autocorrelation of the particle orientation
\(\angles{\hat n(t) \cdot \hat n(0)}\), as shown in
figure~\ref{fig:correlations}(a). We find a nearly exponential decay
with a small persistence at short time. The ABP model, including the
short autocorrelation time \(\tau\), predicts
\(e^{-D_R \left[t - \tau \left(1 - e^{-t/\tau} \right)\right]}\). The
parameter \(\tau\) is held fixed in all subsequent fits at a value
determined from all six particle configurations; this is a short
time-scale, consistent with the autocorrelation in
figure~\ref{fig:velocity}(c). This leaves the exponential decay rate
\(D_R\) as the only free parameter in fitting each particle
configuration.

Second, as shown in figure~\ref{fig:correlations}(b), to characterize
the self-propelled motion we measure the time-dependent correlation
function \(\angles{\vec r(t) \cdot \hat n(0)}\), which is the
displacement, after lag time \(t\), projected along the initial
orientation. As one may anticipate, this initially grows linearly with a
constant speed~\(v_0\) for a duration set by the rotational timescale
\(1/D_R\). Eventually the displacement saturates at a characteristic
length scale, which defines a persistence length \(\ell_p\). After this,
the particle has lost its initial orientation and diffuses
isotropically. The ABP model predicts an antisymmetric function
\(\ell_p e^{\tau D_R}\left(1 - e^{-D_R|t|}\right)\operatorname{sign}(t)\),
which grows linearly at short times with slope \(v_0\) then
asymptotically approaches the persistence length
\(\ell_p = \frac{v_0}{D_R}\). The model describes the data with the
exception of a small but systematic asymmetry that may be due to finite
system size; we therefore fit to the anti-symmetric part of our data.
The fit is a single-parameter fit for the persistence length~\(\ell_p\).

Third, we compute the mean squared displacement
\(\langle[\vec r(t) - \vec r(0)]^2\rangle\), as shown in
figure~\ref{fig:correlations}(c). The ABP prediction for mean squared
displacement is
\(2 \ell_p^2 e^{\tau D_R} \left(D_R t - 1 + e^{-D_R t}\right) + 4 D_T t\).
There are three distinct stages of motion: (i) at times shorter than
\(D_T/v_0^2\), translational diffusion dominates and we expect linear
growth with constant \(D_T\); (ii) later the self-propulsion dominates
and displacement grows ballistically with speed \(v_0\) until (iii) the
orientation diffuses at the timescale of \(1/D_R\) and the displacement
grows linearly again with effective diffusion constant of
\(\ell_p^2 e^{\tau D_R} D_R/2 + D_T\). The model describes the data well
over all of these stages. The earliest timescale, of translational
diffusion, in our experiment is pushed to an extremely short time,
i.e.,~\(D_T/v_0^2 \lesssim 1/f\). Therefore \(D_T\) is not reliably
determined by the fits, and we plot a two-parameter fit with \(\ell_p\)
varied as well. Alongside the total mean squared displacement, we show
the longitudinal and transverse body-frame
components~\citep{supplementary}.

\subsection{Comparing parameters between short- and long-time
measurements}\label{comparing-parameters-between-short--and-long-time-measurements}

\emph{Noise vs.~diffusion and mobility.---} Heretofore, we have
extracted values for \(D_R\), \(\ell_p\), and \(D_T\) via two distinct
methods: from short-time noise statistics and from fitting correlation
functions from the ABP model to the long-time motion of particles. We
have shown this process for one experimental configuration in
figures~\ref{fig:velocity} and~\ref{fig:correlations}. We summarize the
results from six configurations in figure~\ref{fig:parametric} by
checking for consistency between the values extracted from the two
methods. We find that the values agree well, lying close to the line of
slope one in each panel. The agreement is consistent using values
extracted from the single-parameter sequence of fits (red symbols), and
is also robust to fitting with two free parameters (blue symbols). The
assumption of uncorrelated rotational noise yields a poor match between
the two values of \(D_R\) (open symbols). The weakest agreement is for
\(D_T\), which is unsurprising as the early-time diffusive motion barely
appears in our experimental time-window.

\section{Conclusion}\label{conclusion}

Our results demonstrate the commonality between vibrated granular media
and other active matter. Here too, the long-term motion can be
constructed based on independent contributions from noise and mobility,
despite their shared physical origin. The noise has only short temporal
and spatial correlations, and can be reasonably described by a Gaussian
distribution. These observations justify the assumptions of the ABP
model. Further, fitting the model to observed dynamics can reliably
extract parameters of motion. Our results validate the use of this
single-particle description to study collective effects in granular
systems. More generally, we suggest that the sequence of data analysis
laid out in this article is applicable to other active matter, beyond
vibrated granular media.

\subsection{Acknowledgements}\label{acknowledgements}

We gratefully acknowledge funding from NSF-DMR 1506750~and NSF-REU
1359191 (LW, SS, NM); and NSF-DMR 1149266, NSF-MRSEC 1420382, and
IGERT-DGE 1068620 (CW, AB).

\bibliography{noise.bib}

\begin{thebibliography}{28}
\bibitem [{sup()}]{supplementary}%
  \BibitemOpen
  \href@noop {} {}\bibinfo {note} {See {S}upplemental {M}aterial at [{URL} will
  be inserted by publisher] for videos of the experiment and for details of the
  model and data analysis}\BibitemShut {NoStop}%
\end{thebibliography}

\end{document}


\title{Supplementary Information for: Noise and diffusion of a vibrated
self-propelled granular particle}

\author{Lee Walsh}
\email{lawalsh@physics.umass.edu}
\affiliation{Department of Physics, University of Massachusetts, Amherst, USA}

\author{Caleb G.~Wagner}
\affiliation{Martin Fisher School of Physics, Brandeis University, Waltham, MA, USA}

\author{Sarah Schlossberg}
\affiliation{Department of Physics, University of Massachusetts, Amherst, USA}

\author{Christopher Olson}
\affiliation{Department of Physics, University of Massachusetts, Amherst, USA}

\author{Aparna Baskaran}
\affiliation{Martin Fisher School of Physics, Brandeis University, Waltham, MA, USA}

\author{Narayanan Menon}
\email{menon@physics.umass.edu}
\affiliation{Department of Physics, University of Massachusetts, Amherst, USA}

\date{\today}
\maketitle

\section{Theoretical model and
predictions}\label{theoretical-model-and-predictions}

\subsection{Model}\label{model}

An active Brownian particle is parametrized in terms of its position
\(\vec r(t)\) and orientation \(\theta(t)\), which evolve according to
the following set of stochastic differential equations:
\begin{align}
\dot{\vec r}(t) &= v_0 \hat n(t) + \vec \eta(t) \label{eq:abp_T} \\
\dot{\theta}(t) &= \xi(t).
\label{eq:abp_R}
\end{align}
Here \(v_0\) is the magnitude of the self-propulsion velocity, which
points in the direction of the unit vector
\(\hat n(t) = {\cos \theta \choose \sin \theta}\); and the
\(\vec \eta(t)\) and \(\xi(t)\) are stochastic terms that indicate
Gaussian white noise, with correlations
\begin{align}
\angles{\vec \eta_\alpha(t) \, \vec \eta_\beta(t')} &= 2 D_T \delta(t-t') \delta_{\alpha\beta} \label{eq:noise_T} \\
\angles{\xi(t) \, \xi(t')} &= 2 D_R \delta(t - t').
\label{eq:noise_R_white}
\end{align}
In our case, we modify the above standard description to allow for the
temporal correlations in the angular velocity \(\xi(t)\). In fact, there
is a unique way of doing so, provided we retain the original assumptions
that \(\xi(t)\) is Markov, Gaussian, and temporally homogeneous (that
is, neglecting transient contributions from the initial distribution of
\(\xi(0)\))~\citep{doob_brownian_1942}. Under these assumptions, the
correlation is necessarily given as:
\begin{equation}
\angles{\xi(t) \, \xi(t')} = \frac{D_R}{\tau} e^{-|t-t'| / \tau},
\label{eq:noise_R}
\end{equation}
such that \(\tau\) is the autocorrelation time that we obtain from the
experimental data. In summary, Eqs.~(\ref{eq:abp_T}, \ref{eq:abp_R},
\ref{eq:noise_T}, and \ref{eq:noise_R}) constitute our theoretical
model.

\subsection{Correlation functions}\label{correlation-functions}

\emph{Orientation autocorrelation.---} We first study the trajectory of
a single particle with initial orientation \(\theta_0 = \theta(0)\).
Because the rotational noise is Gaussian with zero mean, the following
identity holds:
\begin{align}
\angles{\cos n \theta(t)} &= \cos n\theta_0 \exp\left[-\frac{n^2}{2}\angles{\Delta \theta(t)^2}_c \right], \label{eq:cosine_correlator}
\end{align}
and likewise for \(\angles{\sin n\theta(t)}\). Here \(n\) may be any
integer, and \(\angles{\Delta \theta(t)^2}_c\) is the second cumulant of
the angular displacement. We may obtain this quantity by integration:
\begin{equation}
\angles{\Delta \theta(t)^2}_c
= \angles{\Delta \theta(t)^2}
= \int_0^t \! \int_0^t \angles{\xi(s) \, \xi(s')} ds \, ds'.
\label{eq:second_cumulant}
\end{equation}
Substituting from Eq.~(\ref{eq:noise_R}), we find
\begin{align}
\angles{\Delta \theta(t)^2}_c &= \frac{D_R }{\tau} \int_0^t \left[ \int_0^{s'} e^{-(s' - s) / \tau} ds + \int_{s'}^t  e^{-(s - s') / \tau} ds \right] ds' \\
&= 2D_{R}t-2D_{R}\tau \left( 1-e^{-t / \tau}\right).
\end{align}
Returning to Eq.~\ref{eq:cosine_correlator}, and averaging over initial
orientations \(\theta_0\), we obtain the correlator
\begin{align}
\angles{\cos \theta \cos \theta_0} &= \frac{1}{2} \exp\left[-D_{R}t+D_{R}\tau \left( 1-e^{-t / \tau}\right)\right],
\end{align}
and likewise for \(\angles{\sin \theta \sin \theta_0}\). In terms of
\(\hat n\), we have:
\begin{equation}
\angles{\hat{n}_\alpha (t) \, \hat{n}_\beta(0)} = \frac{\delta_{\alpha \beta}}{2}  \exp\left[-D_{R}t+D_{R}\tau \left( 1-e^{-t / \tau}\right)\right].
\label{eq:nn_corr}
\end{equation}

\emph{Displacement--orientation correlation.---} These results
immediately enable us to solve for the displacement--orientation
correlation:
\begin{align}
\angles{\vec r_\alpha(t)} &= \int_0^t v_0 \angles{\cos\theta(s)} ds + \int_0^t \angles{\vec \eta(s)} ds \\
&= \int_0^t v_0 \cos \theta_0  \exp\left[-D_{R}s+D_{R}\tau \left( 1-e^{-s / \tau}\right)\right] ds + 0.
\end{align}
Averaging over initial orientations, we obtain
\begin{equation}
\angles{\vec r_\alpha(t) \, \hat{n}_\beta(0)} = \delta_{\alpha \beta} \frac{v_0}{2} \int_0^t  \exp\left[-D_{R}s+D_{R}\tau \left( 1-e^{-s / \tau}\right)\right] ds.
\end{equation}
If \(\tau\) is small, we may throw away the double exponential and
evaluate the integral to get
\begin{align}
\angles{\vec r_\alpha(t) \, \hat{n}_\beta(0)} &\simeq \frac{v_0}{2 D_R} e^{D_R \tau} \left(1 - e^{-D_R t} \right) \delta_{\alpha \beta}
\label{eq:rn_corr}
\end{align}

\emph{Mean squared displacement.---} For mean squared displacement, we
again integrate to get
\begin{equation}
\angles{\left[x(t) - x(0)\right]^2} =
v_0^2 \int_0^t \! \int_0^t \angles{\cos \theta(s) \cos \theta(s')} ds \, ds'
+
\int_0^t \! \int_0^t \angles{\vec \eta(s) \, \vec \eta(s')} ds \, ds'.
\label{eq:x_msd}
\end{equation}
Note that the cross-terms vanish because \(\cos \theta(t)\) and
\(\vec \eta(t)\) are independent random variables, and
\(\angles{\vec \eta(t)} = 0\). The latter integral is straightforward,
giving \(2D_T t\). We evaluate the first by the following reasoning from
probability theory. Supposing \(s > s'\), we write
\begin{equation}
\angles{\cos \theta(s) \cos \theta(s')} = \angles{\cos \theta(s') \angles{\cos \theta(s) | \cos \theta(s')}}
\end{equation}
where \(\langle \cos \theta(s) | \cos \theta(s') \rangle\) denotes the
average of \(\cos \theta(s)\) given the (sharp) initial value of
\(\cos \theta(s')\). This quantity may be evaluated using the results of
the previous sections, as follows:
\begin{align}
\angles{\cos \theta(s) | \cos \theta(s')} &= \cos \theta(s') \exp\left[-D_{R}(s-s')+D_{R}\tau \left( 1-e^{-(s - s') / \tau}\right)\right];\\
\angles{\cos \theta(s) \cos \theta(s')} &= \angles{\cos^2 \theta(s')}
\exp\left[-D_{R}(s-s')+D_{R}\tau \left( 1-e^{-(s-s') / \tau}\right)\right];\\
\angles{\cos^2 \theta(s')} &=
\frac{1}{2} + \frac{1}{2}\angles{\cos 2\theta(s')} \\
&= \frac{1}{2} + \frac{1}{2} \cos 2\theta_0 \exp \left[ - 4 D_{R}s'+ 4D_{R}\tau \left( 1-e^{-s' / \tau}\right) \right].
\end{align}
The analogous expression for \(s' > s\) is obtained by swapping \(s\)
and \(s'\). Now, to render the integral in Eq.~\ref{eq:x_msd} tractable,
we again take the limit in which \(\tau\) is very small and drop the
superexponential piece. Substituting into Eq.~\ref{eq:x_msd} and
averaging over~\(\theta_0\), we obtain
\begin{equation}
\angles{\left[x(t) - x(0)\right]^2} = \left(\frac{v_0}{D_R}\right)^2 e^{\tau D_R} \left( e^{-D_Rt} - 1 + D_R t \right) + 2D_T t
\end{equation}
We sum over dimensions to obtain the total mean squared displacement:
\begin{equation}
\angles{\left[\vec r(t) - \vec r(0)\right]^2} = 2\left(\frac{v_0}{D_R}\right)^2 e^{\tau D_R} \left( e^{-D_Rt} - 1 + D_R t \right) + 4D_T t. \label{eq:rr_corr}
\end{equation}
Finally, we consider the longitudinal~(\(\parallel\)) and
transverse~(\(\perp\)) components of the mean squared displacement, in
the body frame of the particle. To calculate this, we take the
expression for \(\langle \left[x(t) - x(0)\right]^2 \rangle\) prior to
any averaging over \(\theta_0\). The longitudinal component then
corresponds to \(\theta_0 = 0\), and transverse to
\(\theta_0 = \pi / 2\). This gives
\begin{equation}
\mathrm{msd}^\parallel_\perp
=
\ell_p^2 e^{\tau D_R} \left(
D_R t - 1 + e^{-D_R t}
\pm
\frac{1}{12} e^{4 \tau D_r} \left(e^{-4D_R t}- 4 e^{-D_R t} + 3\right)
\right) + 2 D_T t.
\end{equation}

\section{Experiment and analysis}\label{experiment-and-analysis}

\subsection{Experimental
configurations}\label{experimental-configurations}

We refer in the paper to six experimental configurations, which give a
range of motility parameter values. Each configuration is a different
combination of:
\begin{itemize}
\tightlist
\item
  two variations in the particle design, with different slopes of the
  beveled nose: 69° and 73°,
\item
  three vibrational amplitudes: \(10g\), \(15g\), and \(20g\).
\end{itemize}

\subsection{Data analysis}\label{data-analysis}

\emph{Image analysis.---} The position and orientation of a particle in
each video frame is resolved in our Python-based tracking
software~\citep{python, numpy, walsh_ordering_2016}. The position
\((x, y)\) is measured to sub-pixel resolution within the image frame as
the intensity-weighted centroid of the segment~\citep{skimage}
corresponding to the marked particle. The orientation \(\theta\) is the
arctangent of the mean of the displacement vectors from the center to
two corners of the particle.

\emph{Velocity.---} The velocities we report in the analysis are
numerical time-derivatives of the measured position and orientation. To
calculate the derivative, we convolve position with the derivative of a
Gaussian kernel~\citep{scipy}. This gives the velocity effectively
averaged over a time~\(\Delta t = 2 \sqrt 3 \sigma\), where \(\sigma\)
is the Gaussian's standard deviation. The vibration period~\(1/f\) sets
the relevant physical timescale of our experiment, below which we assume
the dynamics are not relevant to our present analysis. Our video frame
rate exceeds the vibration time scale, giving a time step of
\(\delta t = 1/(2.4 f)\) between position measurements. Thus, we choose
\(\sigma\) such that \(\delta t < \Delta t < 1/f\).

\emph{Correlations.---} The correlation functions shown in
figures~2(c,d) and 3(a,b,c) of the main text and predicted above in
Eqs.~\ref{eq:noise_T}, \ref{eq:noise_R}, \ref{eq:nn_corr},
\ref{eq:rn_corr}, and \ref{eq:rr_corr} are calculated from the data as
fast-Fourier-transform convolutions~\citep{scipy} of single-particle
trajectories.

\emph{Fitting.---} Here we detail the process used to determine the
model parameters from fitting the correlation functions (shown in
figure~3 of the main text). The three correlation functions
(Eqs.~\ref{eq:nn_corr}, \ref{eq:rn_corr}, \ref{eq:rr_corr}) depend on
one, two, and three parameters. Thus, we introduce one new parameter to
each fit in the sequence. In the primary sequence, we fit each function
with a single free parameter, fixing any other parameters to their
values from the previous fit. In some cases, better fits may be obtained
with two free parameters, whose values are generally consistent with
those from the single-parameter fits. In figure~3 of the main text, we
have plotted single-parameter fits for orientation
autocorrelation~(\ref{eq:nn_corr}) and longitudinal
displacement~(\ref{eq:rn_corr}), and the two-parameter fit for mean
squared displacement~(\ref{eq:rr_corr}). In the parameter comparison
(figure~4, main text) we show the parameters from all single-parameter
fits and the two-parameter fit to mean squared displacement.

\subsection{Spatial inter-particle
correlations}\label{spatial-inter-particle-correlations}

\begin{figure}
\includegraphics[width=\columnwidth]{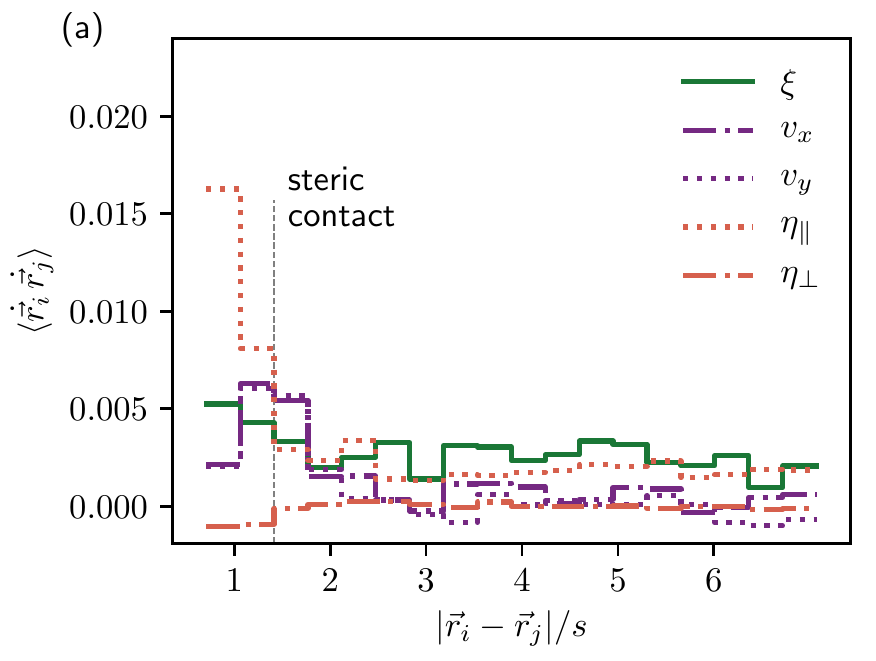}
\caption{
    Spatial inter-particle velocity--velocity correlations in the noise.
    Shown are rotational ($\xi$, in units of $\text{rad}^2 f^2$) and
    translational (in units of $s^2 f^2$) velocity in both the lab
    ($v_x$, $v_y$) and body ($\eta_\parallel$ and $\eta_\perp$) frame,
    plotted as a function of center--center separation between pairs of
    particles. Due to the square shape of the particles, corner--corner
    contact occurs at a separation of $r \approx \sqrt 2 s$ (vertical
    line) while face--face contact is at $r = s$.
}
\label{fig:spatial}
\end{figure}

As described in the main text, despite the highly correlated nature of
the noise source over space as well as time, experimental data clearly
demonstrate that such correlations in the driving force do not generate
significant spatial correlations between particles. Here, we calculate
velocity--velocity radial correlations
\(\angles{\dot{\vec r}_i(t) \, \dot{\vec r}_j(t)}\) as a function of the
pair separation distance \(r\) between two particles at a single point
in time. The average is over time and all particle pairs \(i\) and \(j\)
separated by center--center distance \(r = |\vec r_i - \vec r_j|\). We
calculate these correlations in the rotational and translational
velocity in both the lab and body frames. In several components of the
velocity, correlations due to steric interactions appear at closest
contact (\(r = s\)) and, due to the square particle shape, persist
toward the furthest reach of interactions at the corner--corner contact
\(r \approx \sqrt 2 s\). Beyond this range, correlations vanish in all
velocity components.

\subsection{Moments of the noise
distributions}\label{moments-of-the-noise-distributions}

Notwithstanding any correlations, the ABP model assumes noise to come
from purely Gaussian distributions with zero mean (excepting
longitudinal velocity with mean \(v_0\)), zero skewness, and zero excess
kurtosis. We report in Table~\ref{tab:moments} these moments averaged
over all six experimental configurations. The only substantial deviation
from Gaussian is the skewness of the longitudinal velocity.

\begin{table}[h]
\centering
\begin{tabular}{@{}c c|c|c|c@{}}
\toprule
 & mean & variance & skewness & kurtosis\\
 \midrule
$\xi$ & $-0.02 \pm 0.04$ & $0.069 \pm 0.016$ & $-0.02 \pm  0.12$ & $  4.6 \pm   4.1$\\
$\eta_\parallel$ & $ +0.18 \pm 0.01$ & $ 0.012 \pm 0.002$ & $-0.44 \pm  0.24$ & $ 0.22 \pm  0.23$\\
$\eta_\perp$ & $-0.009 \pm 0.009$ & $0.009 \pm 0.002$ & $+0.09 \pm 0.16$ & $0.05 \pm  0.14$\\
\bottomrule
\end{tabular}
\caption{The mean, variance, skewness, and excess kurtosis of noise, averaged over experimental configurations.}
\label{tab:moments}
\end{table}

\bibliography{supplementary.bib}